\documentstyle[amsfonts,12pt]{article}

\def\half{\textstyle{\frac{1}{2}}}

\def\A{{\frak A}}
\def\P{{\frak P}}

\def\ra{\rightarrow}

\def\s{\hskip.08em}

\def\b{\begin{eqnarray*}}     
\def\e{\end{eqnarray*}}       
\def\bn{\begin{eqnarray*}}     
\def\en{\end{eqnarray*}}       
\def\<{\langle}
\def\>{\rangle}

\def\{{\lbrace}
\def\}{\rbrace}
\def\vp{\varphi}

\begin{document}
\title{Overcoming Nonrenormalizability -- Part 2}
\author{John R. Klauder
\footnote{Electronic mail: klauder@phys.ufl.edu}\\
Departments of Physics and Mathematics\\
University of Florida\\
Gainesville, FL  32611}
\date{}     
\maketitle
\begin{abstract}
The procedures to overcome nonrenormalizability of $\vp^4_n$, $n\ge5$,
quantum field theory models that were presented in a recent paper are
extended to address nonrenormalizability of $\vp^p_3$, $p=8,10,12,\ldots$,
models. The principles involved in these procedures are based on the
hard-core picture of nonrenormalizability. 
\end{abstract}
\subsection*{Introduction}
The present paper may be regarded as an addendum to a recent paper,
 \cite{kla1}, where proposals were advanced to overcome 
nonrenormalizability for $\vp^4_n$ models, quartic self-interacting 
scalar field models with spacetime dimension
$n\ge5$. (Similar techniques were also proposed to overcome triviality 
for the renormalizable but not 
asymptotically free
model $\vp^4_4$.) In this short note we extend the same scheme to advance 
a proposal designed to overcome nonrenormalizability in models
such as $\vp^p_3$, for $n=3$ and powers $p=8,10,12,...\,$, as well.
(If the renormalizable model $\vp^6_3$ is trivial and not asymptotically free, then our proposals 
should 
work for this model as well.)
For background as well as notational questions we urge the reader
to consult \cite{kla1}. 

The philosophy underlying our formulation is based on the hard-core
picture of nonrenormalizable interactions. 
A brief introduction to this viewpoint is offered in \cite{kla1}; a
more detailed discussion appears in \cite{kbook}. As one important consequence
we are led to consider renormalization counterterms that are entirely
different from those suggested by conventional perturbation analyses.
Let us start with an heuristic, motivational discussion.

We work entirely in Euclidean spacetime and assume the theories of interest
arise as suitable continuum and infinite volume limits of a lattice model 
formulated on a
large but finite cubic lattice with a dimensionless lattice spacing
$a$ and periodic boundary conditions. From the viewpoint of critical phenomena
the models in question involve multicritical points, and therefore
upper critical dimensions, above which mean field arguments are generally
applicable, depend on the choice of $p$. It is straightforward to show that
several correlation functions of interest for $\vp^p_3$ models 
are given as follows (see, e.g., \cite{fish,zinn,lub}):
 \bn  &&\hskip2.13cm\Sigma_k\s\<\vp_0\vp_k\>\propto a^{-2}\;,  \\ 
      &&\hskip1.7cm\Sigma_k\s\s k^2\<\vp_0\vp_k\>\propto a^{-4}\;,  \\
    &&\hskip-1.54cm\Sigma_{k_2,k_3,\ldots,k_{2r}}\s\<\vp_0\vp_{k_2}
\vp_{k_3}\cdots\vp_{k_{2r}}\>^T
\propto a^{[2p-4r(p-1)]/[p-2]}\;,   \en
for relevant $r$ values of the form $r=1+j(p-2)/2$, $j=0,1,2,...$ ,  
and where $k\in{\mathbb Z}^3$ denotes a lattice site, $T$ denotes the
truncated (or connected) component, and we have assumed all odd-order correlation functions vanish. It is conventional to recast these
expressions into the single combination
  \bn   g_r\equiv -\s\frac{\Sigma_{k_2,k_3,\dots,k_{2r}}
\<\vp_0\vp_{k_2}\vp_{k_3}\cdots\vp_{k_{2r}}\>^T}{[\s\Sigma_k\<\vp_0\vp_k\>\s]^r\s
[\s\Sigma_k\s k^2\<\vp_0\vp_k\>/6\s\Sigma_k\<\vp_0\vp_k\>\s]^{3(r-1)/2}}\;. \en
The expression for $g_r$ is dimensionless, enjoys
rescaling invariance (i.e., $\vp_k\ra S\s\vp_k$, $S>0$, for all $k$), 
and admits
a meaningful continuum limit. In particular, for small $a$ it follows that 
  \bn  g_r\propto a^{(r-1)(p-6)/(p-2)}\;.  \en
Therefore, when $p\ge8$ and $a\ra0$ we find that $g_r\ra0$ for all 
relevant $r\ge p/2$. This behavior --
which is analogous to what one finds for $\vp^4_n$ models when $n\ge5$ 
\cite{book} --  strongly suggests, in the
continuum limit, that the nonrenormalizable 
$\vp^p_3$ models exhibit ``infidelity''.  By infidelity we mean that the resultant quantum theory has a trivial classical limit, clearly differing from the original classical theory, and thus casting doubt on the quantization procedure itself. This result arises because (i) the quantization loses the $\phi^p_3$ interaction (a conclusion supported by renormalization group analysis), and (ii) any surviving interactions, e.g., $\phi^4_3$ and possibly $\phi^6_3$, have been induced and therefore arise from one or more loop contributions. Such terms therefore have $\hbar$-dependent coupling constants. Thus, in the classical limit where $\hbar\ra 0$ all interactions disappear leading to classical triviality. 

The dependence on the lattice spacing $a$ that has 
led to this claim has arisen from summing over the whole lattice
and is based on divergent behavior that emerges near a second-order phase
transition. If, by some procedure, we could {\it simultaneously} 
rescale all correlation functions {\it uniformly} so that
  \bn  \<\vp_{k_1}\vp_{k_2}\cdots\vp_{k_{2r}}\>\propto a^{(p-6)/(p-2)}
\;,\hskip 1cm {\rm all}\,\;r\ge1\;, \en
then, with this modification taken into account, it follows that
$g_r\propto a^0=1$, for all relevant $r$, and the door to fidelity is 
open. To achieve that
uniform rescaling, we closely follow the scheme presented in \cite{kla1}.  

\subsection*{Alternative Lattice Model}
The sought-for generating functional for lattice-space Schwinger functions
$S\{h\}$ will be expressed in the form
     \bn &&\hskip0cm S\{h\}\equiv [\s T\{h\}\s]^{N_R}  \;,  \\
         && T\{h\}\equiv F\s N\int\s e^{\Sigma\s h_k\vp_k a^3\s-\s\A\s-\s\P}
\;\Pi\s d\vp_k\;.  \en
Let us examine the ingredients in these expressions separately.

The conventional 
probability distribution for a $\vp^p_3$ model is given by
  \bn  D\equiv N\s e^{-\s\A}\;,  \en
where 
  \bn \A\equiv \half\Sigma\s(\vp_{k^*}-\vp_k)^2\s a+\half\s m_o(a)^2
\s\Sigma\s\vp^2_k\s a^3+\lambda(a)\s\Sigma\s\vp^p_k\s a^3\;,  \en
and
  \bn  N^{-1}\equiv \int\s e^{-\s\A}\,\Pi\s d\vp_k\;.  \en
As before, $k\in {\mathbb Z}^3$ denotes a lattice site, and $k^*$
denotes each of the three positive nearest neighbors to $k$.
  
We now modify the distribution $D$ as follows. First, let
  \bn  F\equiv K\s a^{(p-6)/(p-2)}\;,  \en
where $K$ is a positive constant, and focus on the 
region where $F<1$. Next, consider 
   \bn  F\s N\s e^{-\s\A}\;,   \en
which (due to $F$) is no longer a normalized distribution. To restore
normalization, we introduce an auxiliary, nonclassical (proportional to
$\hbar^2$) term given by
   \bn  \P\equiv \half\s A(a)\s\Sigma\s[\s\vp^2_k-B(a)\s]/
[\s\vp^2_k+B(a)\s]^2\s a^3  \en
into the action to yield
   \bn D'\equiv F\s N e^{-\s\A\s-\s\P}\;.  \en
The potential chosen for $\P$ is a regularized version of $A/2\vp^2(x)$, 
which is the only ``pure'' counterterm that introduces no new
dimensional coupling constant for any spacetime dimension. 
The factors $A(a)$ and $B(a)$ that appear in $\P$ are positive and, as discussed below, they
are chosen so that $D'$ is a normalized distribution. 

At this stage all the various correlation functions are small (of order $F$) 
and need to be brought back to normal size. To that end, we raise the generating functional $T\{h\}$ to the power 
$N_R\equiv[\!\![\s a^{-(p-6)/(p-2)}\s]\!\!]$,
where $[\!\![\,\cdot\,]\!\!]$ denotes the integral part of its argument.
The result is a new generating 
functional, $S\{h\}$, all truncated correlation functions of 
which are increased
in magnitude over those of $T\{h\}$ by the factor $N_R$. One may understand this procedure as allowing for the use of reducible sharp-time field operator 
representations, a liberalization well known to be important in advanced quantum field theory studies \cite{str}.

As the final step in our construction we take the
continuum limit $a\ra0$, accompanied by another limit in which the 
lattice size grows to eventually cover all of ${\mathbb R}^3$. 
In the continuum limit we require that $B(a)\ra0$, while $A(a)$ 
may or may not diverge
as $a\ra0$. If $A(a)$ diverges we assume that it diverges at a rate
slower than $a^{-4}$ (see \cite{kla1}); in that case, the discussion 
regarding how the form
of $\P$ has been chosen is identical to the discussion presented 
in \cite{kla1}, and so it is not repeated here. As in \cite{kla1},
after the continuum and infinite volume limits, the final result for $S\{h\}$ 
corresponds to a generalized Poisson distribution \cite{def}.

It is noteworthy that an analogous construction has been carried out rigorously and successfully in {\it one} spacetime dimension, i.e., Euclidean time alone; see
Chap.~10 in \cite{kbook}. This calculation was not motivated by a study of
any of the usual nonrenormalizable theories, but it may be used 
to lend credence to the present proposal for such models when analyzed by similar methods.

\subsubsection*{Approximate evaluation}
In addition, in \cite{kla1}, we presented a crude approximation for 
evaluating the
normalization condition that ensures that $D'$ is a probability distribution.
That argument can also be carried over directly to the present situation.
In this approximate calculation it was assumed that the entire lattice
volume ${\cal V}=(La)^3$, where $L$ denotes the number of lattice points
on one edge of the cubic lattice, is divided into $M$ cells of volume
$v=(\xi a)^3$, where $\xi$ denotes an approximate correlation length. 
For simplicity
in evaluation it was assumed that all field variables
 within a correlation volume $v$
were exactly correlated, while field variables in distinct correlation 
volumes were assumed to be 
entirely uncorrelated. Moreover, in any calculation establishing normalization
of $D'$, one first 
chooses the behavior of $B(a)$ in a
suitable way (see \cite{kla1}), e.g., as 
   \bn B(a)=|\ln(a)|^{-2}\;,  \en
and then determines $A(a)$ in relation to that choice. 
The strong simplifications that were made 
led to a rough, approximate expression for $A(a)$ given by 
   \bn A(a)=(2/M)\s[(p-6)/(p-2)]\s(L\s a)^{-3}\s|\ln(a)|^{-1}\;,  \en
Unfortunately, this result for $A(a)$ is only a leading order estimate, 
which is insensitive to 
important parameters such as $m_0^2$ and $\lambda$. 
More precise determination of $A(a)$ 
would include its dependence on such model parameters, in particular on the
coupling constant $\lambda$.

\subsection*{Pseudofree theory}
It is important to note that the factor $F$ which rescales all the 
correlation functions is {\it independent} of $\lambda$. If we consider the
limit of the interacting theory as $\lambda\ra0^+$, it must be kept in mind 
that the resultant limit will {\it not} be the conventional free theory.
Instead, the limiting theory is what we call the {\it pseudofree} theory 
\cite{kbook,kla1}. The pseudofree theory is therefore the noninteracting
theory to which the interacting theory is continuously connected. Stated
otherwise, the conventional free theory is not even 
continuously connected to the interacting theory,
and therefore perturbation-theoretic generated counterterms are not reliable! 
As a consequence, the pseudofree theory acquires interest in its own
right, and from a computational point of view it would be a good place to
begin because it has one less parameter than the interacting theory. Note
as well that each power $p$ in the $\vp^p_3$ models seems to correspond
to a different pseudofree theory since the parameter $p$ enters into $F$ 
and therefore into $A(a)$
in an apparently significant way. It would be of considerable interest 
if Monte Carlo
methods could be used to satisfy the normalization condition and
thereby to help determine the pseudofree theory for one or more $p$ values.

\subsection*{Quantum Fields}
We expect all the models discussed
in this paper to correspond to quantum field theories after Wick rotation
for the following reasons: The essential requirements to lead to a quantum field theory are Euclidean invariance, reflection positivity, moment growth, and clustering. These conditions are satisfied by the original theory and are not disturbed by an overall scaling ($F$) and an additional local interaction ($\frak P$).
Multiple copies also preserve these properties. So long as there is a 
uniform lower bound on the mass, the continuum and infinite volume limits
should have the desired effect.

\subsection*{Other Models}
Although we have confined our attention in this paper to $\vp^p_3$ models
for values of $p\ge8$, it should be evident that similar methods can be
extended to multicritical points associated with other nonrenormalizable
models such as $\vp^p_n$ whenever $p>2n/(n-2)$ and $n\ge4$. This analysis 
would therefore extend the class of models considered in \cite{kla1}.
The essential changes to major formulas given in this paper would be 
that in this more general case
   \bn g_r\propto a^{(r-1)[n(p-2)-2p]/[p-2]}\;, \en
for all relevant $r$, which would involve a uniform rescaling such that
   \bn \<\vp_{k_1}\vp_{k_2}\cdots\vp_{k_{2r}}\>\propto a^{[n(p-2)-2p]/[p-2]}
\;,\hskip 1cm {\rm all}\,\;r\ge1\;. \en
To obtain this rescaling requires that
  \bn F=K\s a^{[n(p-2)-2p]/[p-2]}\;,  \en
and, correspondingly, that
  \bn  N_R= [\!\![\s a^{-[n(p-2)-2p]/[p-2]}\s]\!\!]\;.  \en
With these changes, the discussion is substantially similar to 
that given in the
present paper, augmented when necessary by the contents of \cite{kla1}.

\subsection*{Dedication}
It is a pleasure to dedicate this article to the 70th birthday of
Elliott Lieb. Elliott is a long-time friend and someone I have admired
for his analytical skills and his remarkable originality for many years.
I wish him a long life, full of happiness and continued quality research.

\subsection*{Acknowledgements}
Thanks are expressed to Sergei Obukhov for helpful comments.
This work was partially supported by NSF Grant 1614503-12.


\begin{thebibliography}{99}

\bibitem{kla1} J.R.~Klauder, ``Overcoming Nonrenormalizability'', Lett. 
Math. Phys. {\bf 63}, 229 (2003); hep-th/0209177.

\bibitem{kbook} J.R. Klauder, {\it Beyond Conventional Quantization}, 
(Cambridge University Press, Cambridge, 2000).

\bibitem{fish} M.~Fisher, Rep. Prog. Phys. {\bf 30}, 615 (1967).

\bibitem{zinn} J.~Zinn-Justin, {\it Quantum Field Theory and Critical
Phenomena} 2nd. Ed., (Clarendon Press, Oxford, 1993).

\bibitem{lub} P.M.~Chaikin and T.C.~Lubensky, {\it Principles of Condensed 
Matter Physics}, (Cambridge University Press, Cambridge, 1995).

\bibitem{book} R.~Fern\'andes, J.~Fr\"ohlich, and A.~Sokal, {\it Random
Walks, Critical Phenomena, and Triviality in Quantum Field Theory}, 
(Springer-Verlag, New York, 1992).

\bibitem{str} R.W.~Streater and A.S.~Wightman, {\it PCT, Spin and Statistics, and All That}, (W.A.~Benjamin, New York, 1964); R.~Haag, {\it Local Quantum Physics}, (Springer-Verlag, Berlin, 1992). 

\bibitem{def} B.~De Finetti, {\it Theory of Probability, Volume 2}, 
(John Wiley \& Sons, Bristol, 1975).
\end{thebibliography}
\end{document}